\pdfoutput=1
\documentclass[12pt]{article}
\usepackage[margin=1in]{geometry}
\usepackage{amsmath}
\usepackage{titling}
\usepackage{blindtext}
\usepackage{pgfplots}

\usepackage{comment}
\usepackage[colorlinks=true,linkcolor=blue,allcolors=blue]{hyperref}
\usepackage{url}
\usepackage{setspace}
\usepackage{authblk}
\pgfplotsset{compat=1.14}

\begin{document}

\title{A data-driven MHD model of the weakly-ionized chromosphere}


\author[1]{M. S. Yalim}
\affil[1]{Center for Space Plasma and Aeronomic Research, The University of Alabama in Huntsville, Huntsville, AL 35805, USA}

\author[1]{A. Prasad}

\author[1,2]{N. V. Pogorelov}
\affil[2]{Department of Space Science, The University of Alabama in Huntsville, Huntsville, AL 35805, USA}

\author[1,2]{G. P. Zank}

\author[1,2]{Q. Hu}

\setcounter{Maxaffil}{0}
\renewcommand\Affilfont{\itshape\small}
\date{}    
\begin{titlingpage}
    \maketitle
    

\begin{abstract}
The physics of the solar chromosphere is complex from both theoretical and modeling perspectives. The plasma temperature from the photosphere to corona increases from $\sim$5,000 K to $\sim$1 million K over a distance of only $\sim$10,000 km from the chromosphere and the transition region. Certain regions of the solar atmosphere have sufficiently low temperature and ionization rates to be considered as weakly-ionized. In particular, this is true at the lower chromosphere. In this paper, we present an overview of our data-driven magnetohydrodynamics model for the weakly-ionized chromosphere and show a benchmark result. It utilizes the Cowling resistivity which is orders of magnitude greater than the Coulomb resistivity. Ohm's law therefore includes anisotropic dissipation. We investigate the effects of the Cowling resistivity on heating and magnetic reconnection in the chromosphere as the flare-producing active region (AR) 11166 evolves. In particular, we analyze a C2.0 flare emerging from AR11166 and find a normalized reconnection rate of 0.12.


\end{abstract}
\end{titlingpage}



\section{Introduction} \label{sec:intro}

Chromosphere is a particularly difficult region to model in the solar atmosphere. Together with the transition region, it can be considered as a transition layer from the photosphere to the corona where the plasma temperature increases from $\sim$5,000 K to $\sim$1 million K over a distance of only $\sim$10,000 km. The low temperature in the photosphere results in an ionization fraction of about $n_{i}/n\approx 10^{-4}$ where 1 corresponds to fully ionized plasma. In the chromosphere, the ionization fraction increases but always remains below 1~\cite{Khomenko16}. The dominant mechanism for ionization in the chromosphere is photoionization with the rate for hydrogen $\approx$ 0.014 $\mathrm{s}^{-1}$, which is orders of magnitude greater than the ionization rate due to electron collisions of $\approx 7.8\times10^{-5}$ $\mathrm{s}^{-1}$~\cite{PM98}. Therefore, in order to model the chromosphere properly, the physics of weakly-ionized plasmas should be taken into account. 

Chromosphere models can be classified as part of either global solar atmosphere models (e.g.,~\cite{Holst14,Lionello09}) or local models that simulate the evolution of active regions (ARs). In the latter category (e.g.,~\cite{LA06,Gudiksen11,Rempel17}), there are flux emergence models that extend through the upper convection zone, photosphere, chromosphere, and transition region into the corona. These are local models that include the physical processes to model the weakly-ionized plasma in the chromosphere.

In this paper, we present a data-driven magnetohydrodynamics (MHD) model for the weakly-ionized chromosphere. It is driven by photospheric vector magnetogram data from the Helioseismic and Magnetic Imager (HMI)~\cite{Schou12} onboard the Solar Dynamics Observatory (\emph{SDO})~\cite{Pesnell12}, hence we do not consider the upper convection zone unlike the flux emergence models.~\cite{Leake17} indicates that the data-driven MHD simulations of AR evolution give accurate results when driven by magnetogram data with a cadence of 12 minutes. This result is quite promising for the utilization of HMI vector magnetogram data with a cadence of 12 minutes to drive such simulations. This will give us the possibility to eventually couple our chromosphere model with our data-driven MHD model for global solar corona~\cite{Yalim17}.

We take the partial ionization effects in the chromospheric plasma into account by including the Cowling resistivity.~\cite{Cowling57} showed that the electrical conductivity (the Cowling conductivity) of a non-stationary plasma can be significantly decreased owing to ion acceleration by Ampere's force. Collisions between ion and neutral particles become very effective because of the high ion velocities. As a result, the magnetic flux is not conserved and the rate of magnetic reconnection might be considerably increased due to Joule's (Ohmic) dissipation~\cite{TS10}.

From the perspective of energy balance in the weakly-ionized chromosphere, Cowling resistivity leads to additional dissipation of currents perpendicular to the magnetic field resulting in Joule heating that is several orders of magnitude larger compared to the fully ionized plasma.

Section~\ref{model} gives an overview of our data-driven MHD model for the weakly-ionized chromosphere and presents the benchmark result. In section~\ref{cowling}, we focus on the Cowling resistivity and discuss its effects on heating and magnetic reconnection in the chromosphere based on an analysis using extreme-ultraviolet (EUV) channel 171 \AA~ images from the Atmospheric Imaging Assembly (AIA)~\cite{Lemen12} onboard \emph{SDO} and photospheric vector magnetograms from \emph{SDO}/HMI. In particular, we follow the evolution of AR11166, and its effect on magnetic reconnection and the formation of a C2.0 flare. Finally, section~\ref{conc} presents our conclusions.

\section{Data-driven MHD model for the weakly-ionized chromosphere}
\label{model}
In this section, we present an overview of our data-driven MHD model in terms of the modeling framework, governing equations, initial and boundary conditions, and numerical methods, respectively. We also show a result from a benchmark case obtained with our model.

\subsection{Modeling framework}
\label{framework}
The modeling framework for our MHD simulations is Multi-Scale Fluid-Kinetic Simulation Suite (MS-FLUKSS)~\cite{Pogorelov14}. It allows multi-fluid and coupled MHD plasma / kinetic neutral atoms simulations of partially-ionized plasma. It can also treat pickup ions (e.g.,~\cite{Adhikari15,Pogorelov16}). MS-FLUKSS consists of subroutines implemented on both Cartesian and spherical meshes to solve MHD, Euler, and kinetic Boltzmann equations. It is built upon the Chombo adaptive mesh refinement (AMR) framework~\cite{Colella07} developed at the Lawrence Berkeley National Laboratory. MS-FLUKSS utilizes parallel HDF5 library to store and manage data. It has excellent scalability up to 150,000 cores on major national supercomputers such as ORNL's Jaguar and Titan, NCSA’s Blue Waters, NASA’s Pleiades, and conventional Linux clusters.

MS-FLUKSS is very ﬂexible with respect to specifying different time-dependent BCs (e.g.,~\cite{Yalim17,Kim16,Pogorelov13}). It reads input data sets (observational or numerically obtained, including those from other numerical codes) and interpolates them to chosen spatial and temporal grids.

\subsection{Governing equations}
\label{goveq}
Our model of the weakly-ionized chromosphere is based on the extended resistive MHD equations for hydrogen plasma for any degree of ionization~\cite{LA06}:
\begin{equation}
\label{eqChrom1}
\resizebox{.94\hsize}{!}{$\frac{\partial}{\partial t}
\left(
\begin{array}{c}
\rho  \\
\rho \mathbf{v} \\
\mathbf{B} \\
E
\end{array}
\right) +
\mathbf{\nabla}\cdot
\left(
\begin{array}{c}
\rho\mathbf{v} \\
\rho\mathbf{v}\mathbf{v} + \mathbf{I}(p+\frac{B^2}{8\pi})-\frac{\mathbf{B}\mathbf{B}}{4\pi} \\
\mathbf{v}\mathbf{B}-\mathbf{B}\mathbf{v} \\
(E+p+\frac{B^2}{8\pi})\mathbf{v}-\frac{\mathbf{B}}{4\pi}(\mathbf{v}\cdot\mathbf{B})
\end{array}
\right) -
\left(
\begin{array}{c}
0 \\
\mathbf{\nabla}\cdot\mathbf{\tau} \\
-\mathbf{\nabla}\times\eta\mathbf{J_{\parallel}}-\mathbf{\nabla}\times\eta_{C}\mathbf{J_{\perp}} \\
\mathbf{\nabla}\cdot (\mathbf{B}\times\eta\mathbf{J_{\parallel}})+\mathbf{\nabla}\cdot (\mathbf{B}\times\eta_{C}\mathbf{J_{\perp}})
\end{array}
\right)
=
\left(
\begin{array}{c}
0 \\
\rho\mathbf{g} \\
0 \\
\rho(\mathbf{v}\cdot\mathbf{g})+S_{NA}
\end{array}
\right),$}
\end{equation}
\noindent where $\tau_{ij}=\nu\big[\frac{1}{2}\big(\frac{\partial v_{i}}{\partial x_{j}}+\frac{\partial v_{j}}{\partial x_{i}}\big)-\frac{1}{3}\delta_{ij}\mathbf{\nabla}\cdot\mathbf{v}\big]$ is the viscous stress tensor, $\eta$ is the Coulomb resistivity, $\eta_{C}$ is the Cowling resistivity, $\mathbf{J_{\parallel}}$ and $\mathbf{J_{\perp}}$ are the components of current density parallel and perpendicular to the magnetic field, and $S_{NA}$ is the combination of non-adiabatic source terms corresponding to viscous heating, shock heating, thermal conduction, radiative transfer, and coronal heating.

To evaluate the expression for the Cowling resistivity, $\eta_{C}$, an estimate for the neutral fraction $\xi_{n}$ is required as a function of density and temperature (to be described below). 

Following the method of~\cite{DePontieu99} an electro-neutral hydrogen plasma is assumed. The solar chromosphere is not in LTE, hence a simple one-level model for the hydrogen atom is inadequate for these conditions~\cite{PT59}. To calculate ionization degrees in a non-LTE situation requires the solution of the radiative transfer and statistical equilibrium equations. These are very time consuming to calculate. For this reason, approximations of non-LTE effects on hydrogen ionization have been developed. Accordingly, a two-level model is used for the hydrogen atom, as this provides us with a good approximation for hydrogen ionization at chromospheric densities and temperatures~\cite{TA61}. Under this approximation, the ionization equation~\cite{Brown73} is solved assuming that thermal collisional ionization is not important when compared to photoionization~\cite{Ambartsumyan58}. The steady state solution to this equation is given by~\cite{TA61} (i.e., the modified Saha equation for non-LTE chromosphere):
\begin{equation}
\label{eqChrom3}
\frac{n_{i}^2}{n_{n}}=\frac{f(T)}{b(T)}, 
\end{equation}
with
\begin{equation}
\label{eqChrom4}
f(T)=\frac{(2\pi m_{e}k_{B}T)^{3/2}}{h^3}\mathrm{exp}\Bigl(-\frac{X_{i}}{k_{B}T}\Bigr), 
\end{equation}
and 
\begin{equation}
\label{eqChrom5}
b(T)=\frac{T}{w T_{R}}\mathrm{exp}\left[\frac{X_{i}}{4 k_{B}T}\left(\frac{T}{T_{R}}-1\right)\right],
\end{equation}
where $k_{B}$ is the Boltzmann constant, $h$ is Planck's constant, $X_{i}$ is the ionization energy of the hydrogen atom, $T_{R}$ is the temperature of the photospheric radiation field and $w$ is its dilution factor.

Using Eq.~\ref{eqChrom3}, the ratio of the number density of neutrals to ions is given by 
\begin{equation}
\label{eqChrom6}
r=\frac{n_{n}}{n_{i}}=\frac{1}{2}\left(-1+\sqrt{\left(1+\frac{4\rho/ m_{i}}{n_{i}^2/ n_{n}}\right)}\right),
\end{equation}
and $\xi_{n}=\frac{\rho_{n}}{\rho}=\frac{r}{1+r}$ is the neutral fraction for a hydrogen plasma (where $m_{i}=m_{n}$). The approximation $\rho\approx m_{i}n_{i}+m_{n}n_{n}=m_{i}(n_{i}+n_{n})$ is used as the mass of the electron is small compared with the proton/neutron.

The relation between the Cowling and Coulomb resistivities is
\begin{equation}
\label{eqChrom7}
\frac{\xi_{n}^2 B_{0}^2}{\alpha_{n}}=\eta_{C}-\eta,
\end{equation}
where $B_{0}$ is the magnetic field strength and $\alpha_{n}=m_{e}n_{e}\nu_{en}^\prime+m_{i}n_{i}\nu_{in}^\prime$ with $\nu_{en}^\prime$ and $\nu_{in}^\prime$ defined as the effective collisional frequencies of electrons and ions with neutrals, respectively. Assuming that the chromospheric plasma is entirely composed of hydrogen, 
\begin{equation}
\label{eqChrom8}
\alpha_{n}=\frac{1}{2}\xi_{n}\big(1-\xi_{n}\big)\frac{\rho^{2}}{m_{n}}\sqrt{\frac{16 k_{B}T}{\pi m_{i}}}\Sigma_{in}, 
\end{equation}
where $\Sigma_{in}$ is the ion-neutral cross-section for a hydrogen plasma.

The Coulomb resistivity is computed from 
\begin{equation}
\label{eqChrom9}
\eta=\frac{m_{e}\big(\nu_{ei}^\prime+\nu_{en}^\prime\big)}{n_{e}e^{2}},
\end{equation}
where $e$ is the charge of an electron, $\nu_{en}^\prime$ and $\nu_{ei}^\prime$ are the effective collisional frequencies given by 
\begin{equation}
\label{eqChrom10}
\nu_{en}^\prime=\frac{m_{n}}{m_{n}+m_{e}}\nu_{en},
\end{equation}
and
\begin{equation}
\label{eqChrom11}
\nu_{ei}^\prime=\frac{m_{i}}{m_{i}+m_{e}}\nu_{ei}.
\end{equation} 
Following the example of~\cite{Spitzer62}, the collisional frequencies of electrons with ions and neutrals are estimated by 
\begin{equation}
\label{eqChrom12}
\nu_{en}=n_{n}\sqrt{\frac{8 k_{B} T}{\pi m_{en}}}\Sigma_{en},
\end{equation}
and 
\begin{equation}
\label{eqChrom13}
\nu_{ei}=3.7\times10^{-6}\frac{n_{i}(\mathrm{ln}\Lambda) Z^{2}}{T^{3/2}}, 
\end{equation}
where $m_{en}=\frac{m_{e} m_{n}}{m_{e}+m_{n}}$, $\Sigma_{en}$ is the electron-neutral cross-section for a hydrogen plasma, $n_{n}=\frac{\xi_{n}m_{i}n_{i}}{m_{n}\big(1-\xi_{n}\big)}$ is the neutral number density, $Z$ is the atomic number of hydrogen, and $\mathrm{ln}\Lambda$ is the Coulomb logarithm tabulated in~\cite{Spitzer62}.

\begin{figure}[!http]
\centering
\includegraphics[width=0.485\textwidth]{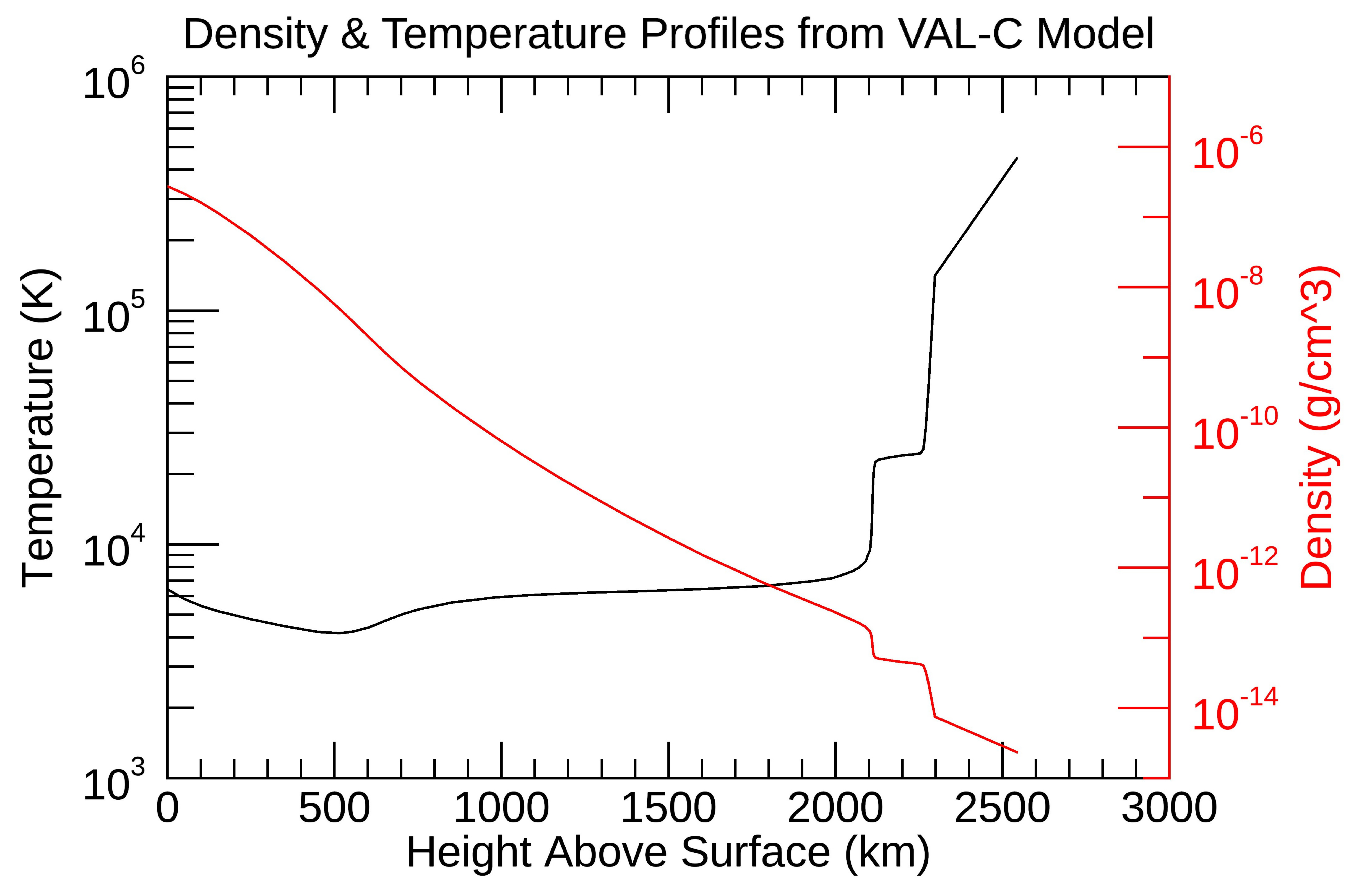}
\includegraphics[width=0.485\textwidth]{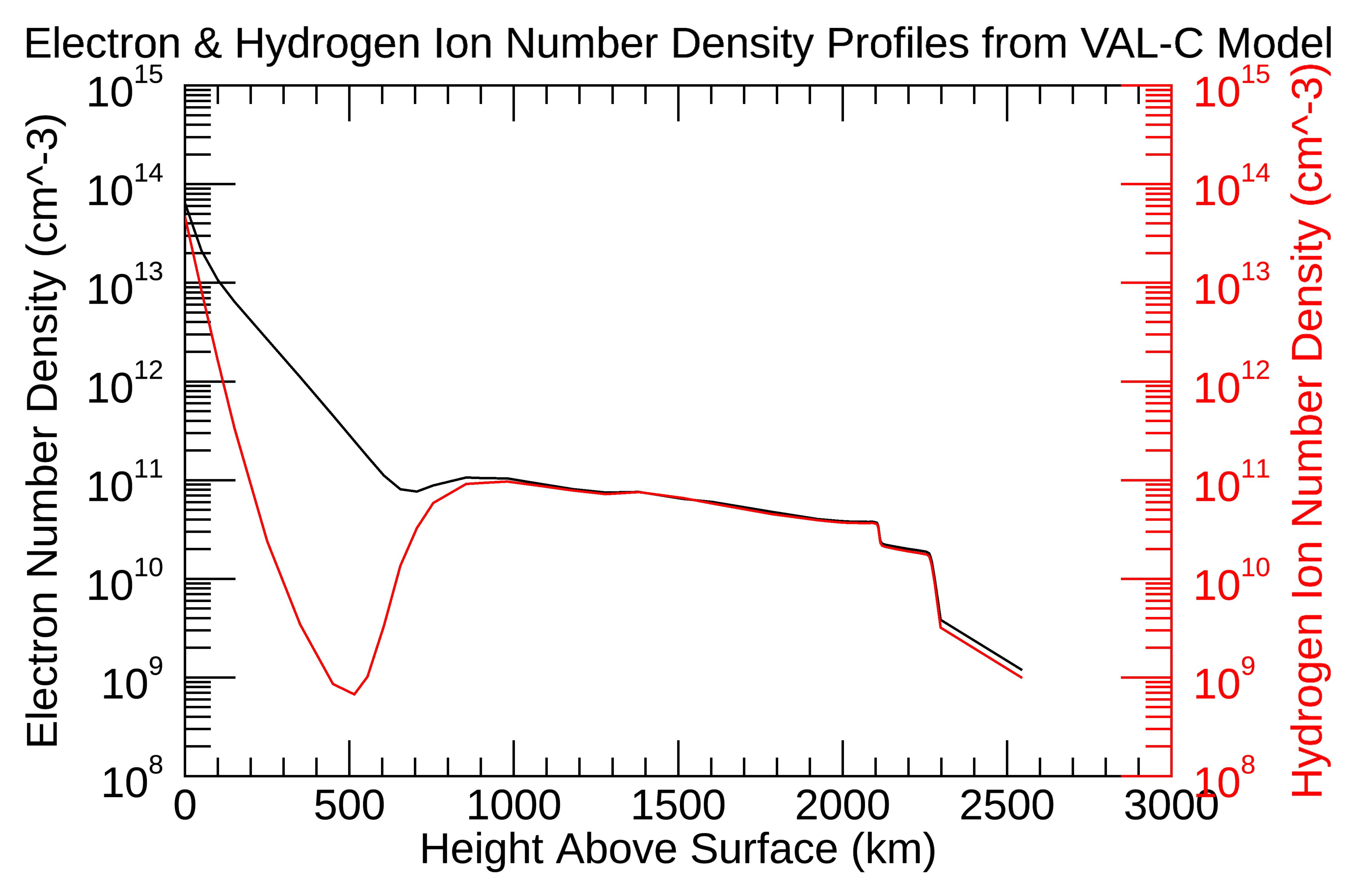}
\caption{(Left) Density (g/$\mathrm{cm^{3}}$) and temperature (K) profiles; (right) electron ($\mathrm{cm^{-3}}$) and hydrogen ion number density ($\mathrm{cm^{-3}}$) profiles in the chromosphere obtained from the VAL-C model~\cite{Vernazza81}.}
\label{fig1}
\end{figure}

To calculate the Coulomb and Cowling resistivities, we need the plasma bulk density and temperature as well as the ion and electron number densities, $n_{i}$ and $n_{e}$, in the chromosphere. These values are tabulated by the VAL-C model~\cite{Vernazza81} based on Skylab observations (see Figure~\ref{fig1}). We compute the magnetic field from the non-force-free-field (NFFF) extrapolation technique~\cite{HD08,Hu10} based on the photospheric vector magnetograms from \emph{SDO}/HMI.

\subsection{Initial and boundary conditions}
\label{icbcs}
The computational domain consists of a box with the bottom and top boundaries located at the photosphere and lower corona, respectively. The box size is chosen to be sufficiently large to cover the AR. For example, for AR11166, which is the AR of interest in section~\ref{cowling}, it is 203$\times$128$\times$128 Mm in the x, y and z directions, respectively where the horizontal plane is indicated by the x and y axes, and z axis is the height from the photosphere. 

The initial solution for magnetic field is computed from the NFFF extrapolation. The chromospheric plasma is assumed to be initially at rest. We specify stratified temperature and density distributions obtained from the VAL-C model between 0-2,500 km height from the photosphere. Above 2,500 km, we assume a normalized stratified temperature distribution as in~\cite{LA06,Jiang16}, and a density distribution calculated from the hydrostatic equilibrium equation and the ideal gas law accordingly. Our computational grid is stratified with height in accordance with the stratified chromospheric plasma parameter distributions in order to resolve the large gradient of the plasma parameters near the photosphere and meanwhile avoid too much computational overhead~\cite{Jiang16}.  

We apply characteristic boundary conditions~\cite{Yalim17,SYP18} inspired from~\cite{Hirsch94} to drive our model by observational data on the photosphere (i.e., HMI vector magnetogram data and horizontal velocity data computed by applying the DAVE4VM method~\cite{Schuck08,Liu13} to vector magnetograms) to solve for a realistic solar chromosphere model. This will allow us to specify sufficient number of mathematically admissible boundary conditions. In the outer and side boundaries, we have the option to specify characteristic or non-reflective boundary conditions.

\subsection{Numerical methods}
\label{nummet}
We discretize the advective fluxes in Eq.~\ref{eqChrom1} using an upwind cell-centered Finite Volume method based on approximate Riemann solvers (i.e., total variation diminishing (TVD) Roe's scheme or Rusanov's scheme). The boundary conditions are specified in the layers of ghost cells located outside the domain adjacent to the boundary cells. The diffusive fluxes are discretized using central discretization. The time derivatives are discretized using explicit time integration schemes implemented in MS-FLUKSS, namely forward Euler, Runge-Kutta 2-step, and Hancock's scheme. In order to increase the convergence speed, we apply local time-stepping.

Since explicit schemes are used for time integration, we determine the iteration timestep $\Delta t$ according to the CFL condition. For this model, the value of $\eta_{C}$ becomes the dominant effect on the condition for stability of the numerical solution to the induction equation as the ratio of $\frac{\eta_{C}}{\eta}$ can be orders of magnitude (see Figure~\ref{fig4} (left)). In order to avoid subjecting the whole scheme to this diffusion timestep condition, the resistive update for a weakly-ionized plasma should be done separately from the main update for ideal MHD by sub-cycling the resistive update inside each ideal step~\cite{LA06}.

Let us recall the $S_{NA}$ term in Eq.~\ref{eqChrom1} which is the combination of non-adiabatic source terms in the energy equation. These terms are difficult to model quantitatively. In order to model the cumulative effects of these terms, we solve a Newton-cooling equation at each timestep in addition to the MHD equations to relax the specific energy density to match the plasma temperature to the initial stratified chromospheric temperature profile given in subsection~\ref{icbcs}. This equation takes the form $\frac{dE}{dt}=-\frac{E-E_{0}(\rho)}{\tau}$ where the specific energy of the plasma is relaxed to its initial state in a time scale $\tau=\big(\frac{\rho}{\rho_{\mathrm{photosphere}}}\big)^{-1.7}$~\cite{GN05}.

\subsection{Benchmark case result}
\label{benchmark}
\begin{figure}[!http]
\centering
\includegraphics[width=0.5\textwidth]{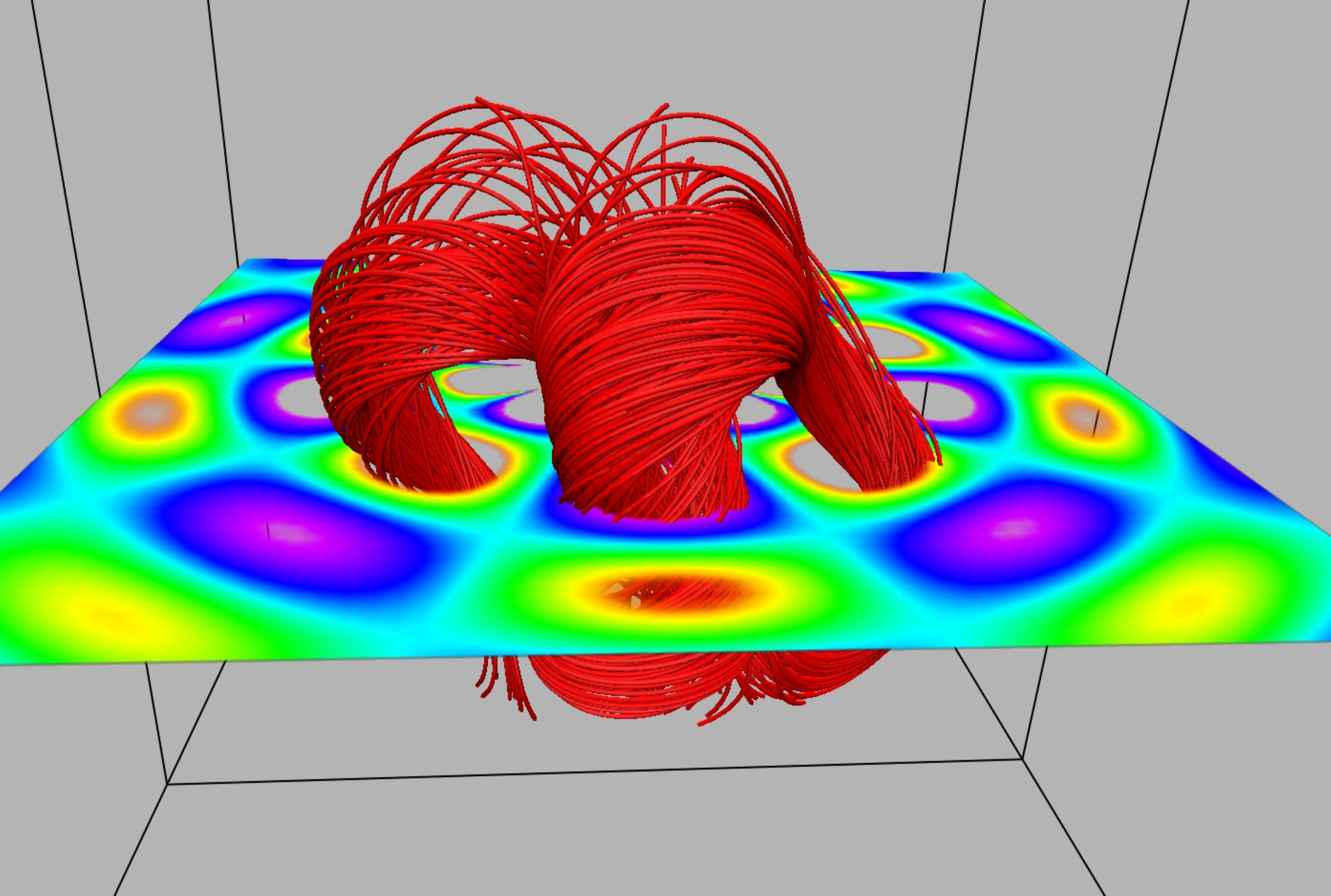}
\caption{Benchmark case: Relaxation of a magnetic flux surface~\cite{PB16} extrapolated by the LFFF technique from an analytical magnetic field data at the bottom boundary that is initially in hydrostatic equilibrium.}
\label{figChromBench}
\end{figure}

In Figure~\ref{figChromBench}, a benchmark result obtained by our chromosphere model shows an analytical magnetic field topology extrapolated by using the linear-force-free-field (LFFF) technique that is initially in hydrostatic equilibrium. The hydrostatic equilibrium and hence the initial extrapolated magnetic field topology is preserved throughout the whole simulation, validating our local simulation setup.

\section{Cowling resistivity in the chromosphere}
\label{cowling}
In this section, we discuss the variation and effects of the Cowling resistivity on the heating and magnetic reconnection in the chromosphere, in particular during the evolution of a flare-producing active region AR11166.

As can be seen in the induction and energy equations in Eq.~\ref{eqChrom1}, the Cowling resistivity dissipates currents perpendicular to the magnetic field while the Coulomb resistivity dissipates currents parallel to it. In addition, Cowling resistivity contributes to heating the chromosphere via the frictional Joule heating term that follows from the generalized Ohm's law according to~\cite{LA06}:
\begin{equation}
\label{eqChrom2}
Q=(\mathbf{E}+(\mathbf{v}\times\mathbf{B}))\cdot\mathbf{j}=\eta J_{\parallel}^2+\eta_{C}J_{\perp}^2.
\end{equation}

\begin{figure}[!http]
\centering
\includegraphics[width=0.49\textwidth]{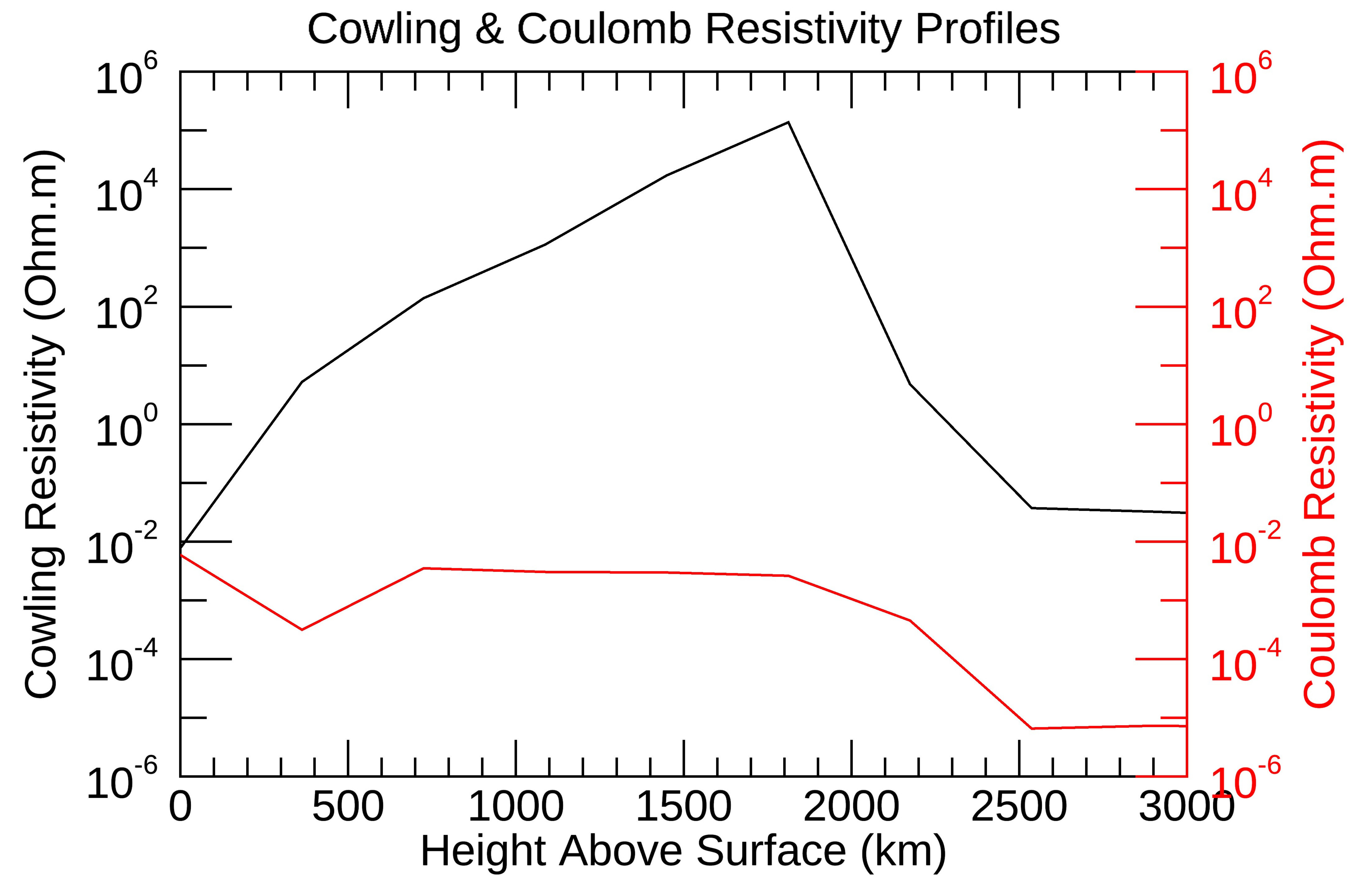}
\includegraphics[width=0.5\textwidth]{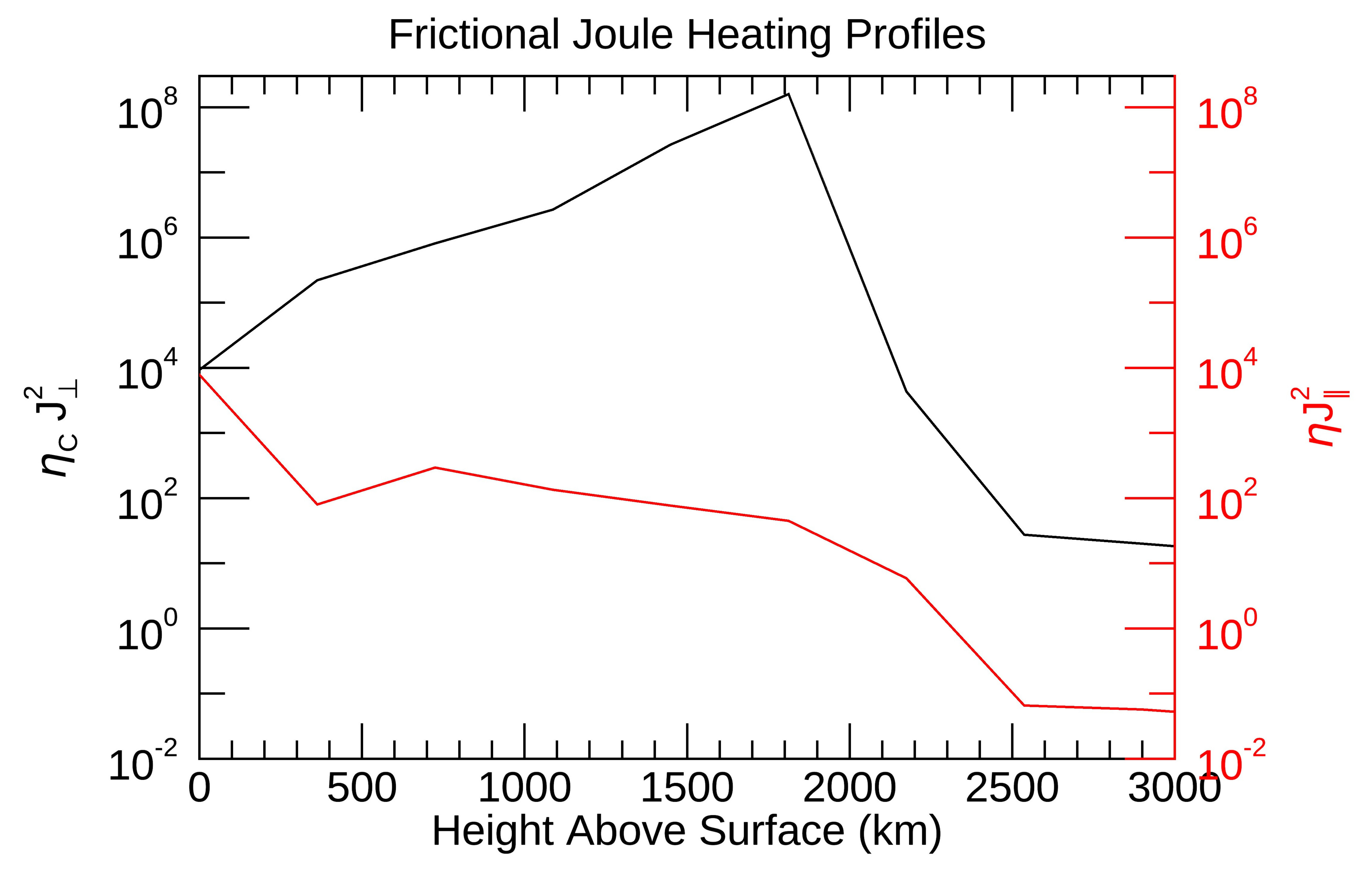}
\includegraphics[width=0.49\textwidth]{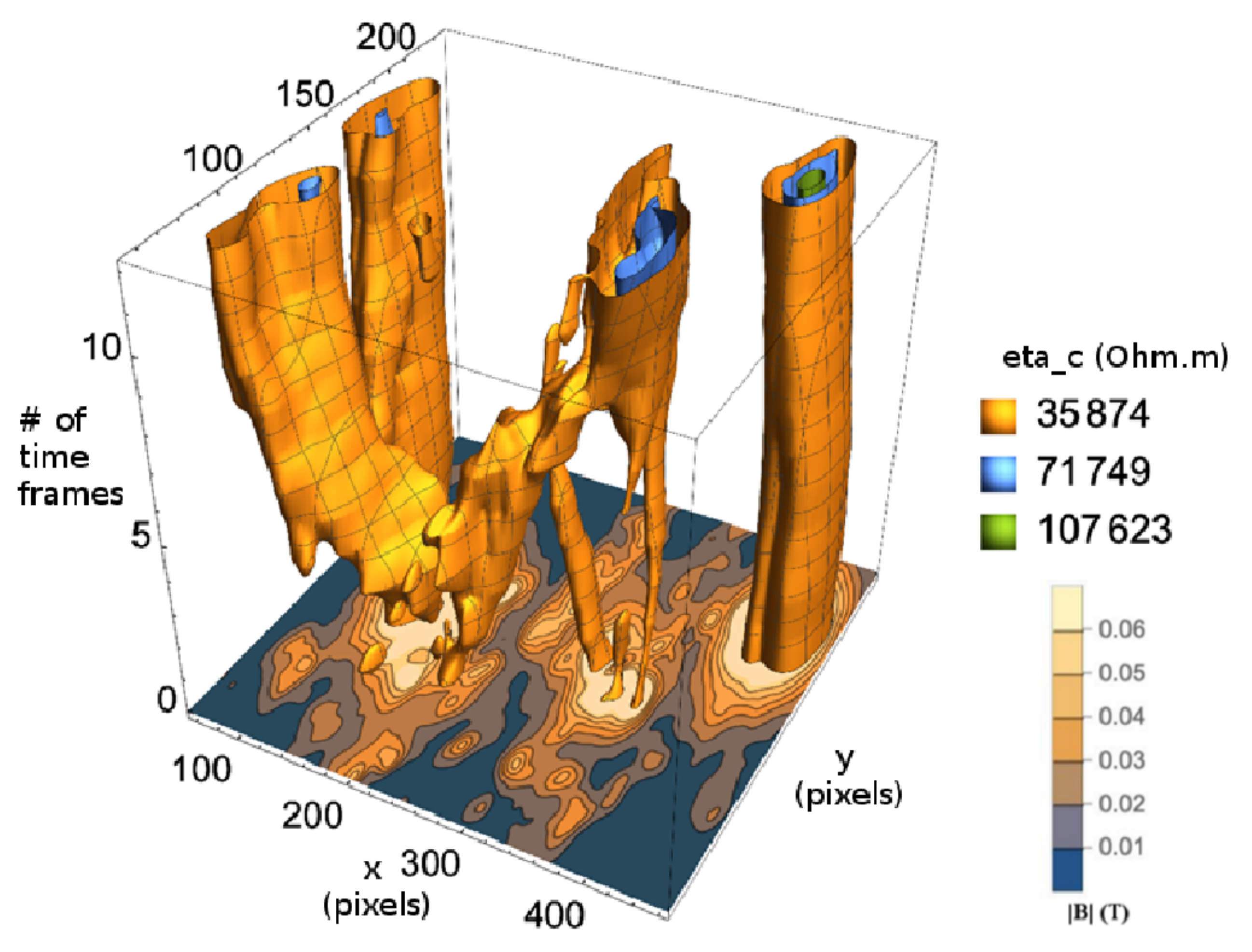}
\caption{Variations of (left) the maximum values of Cowling and Coulomb resistivity profiles with height above the photosphere, and (right) the maximum values of frictional Joule heating profiles (in scaled units) with height above the photosphere for AR11166 at 2011-03-07T06:00:29 UT; (bottom) time-dependent variation of Cowling resistivity at $\sim$1.8 Mm height above the photosphere during the evolution of AR11166. The bottom boundary shows the variation of magnetic field strength at $\sim$1.8 Mm height above the photosphere for AR11166 at 2011-03-07T06:00:29 UT.}
\label{fig4}
\end{figure}

We observe the evolution of AR11166 at 13 timesteps between 2011-03-07T06:00:29 UT and 2011-03-11T06:00:29 UT with a cadence of 8 hours. Figure~\ref{fig4} (left) shows the variations of the maximum values of Cowling and Coulomb resistivity profiles with height at 2011-03-07T06:00:29 UT. Accordingly, the Cowling resistivity is orders of magnitude larger than the Coulomb resistivity in the chromosphere, especially between 1-2 Mm. Figure~\ref{fig4} (right) presents the variation of the maximum values of the frictional Joule heating profiles with height in the chromosphere due to Cowling and Coulomb resistivities. It shows that the chromospheric heating due to the dissipation of currents perpendicular to the magnetic field dominates the heating due to the dissipation of currents parallel to it. This figure demonstrates the significance of Cowling resistivity for chromospheric heating. Figure~\ref{fig4} (bottom) shows the time-dependent variation of Cowling resistivity at $\sim$1.8 Mm height above the photosphere during the evolution of AR11166. The Cowling resistivity distribution follows the AR structure quite well primarily due to its strong dependence on the magnetic field strength (see Eq.~\ref{eqChrom7}). For this reason, its time variation shown in Figure~\ref{fig4} (bottom) can reveal how different structures on the AR evolve in time. Accordingly, the structures at the upper-right and upper-left do not change much as can be deduced from the vertical non-interacting contour structures whereas the other structures interact with each other above the polarity inversion line.

Since the Cowling resistivity is orders of magnitude larger than the Coulomb resistivity in the chromosphere, it can in principle increase the magnetic reconnection rate significantly, and hence play a role in the flare formation, especially in a low-lying 3D null point configuration.

\begin{figure}[!http]
\centering
\includegraphics[width=0.45\textwidth]{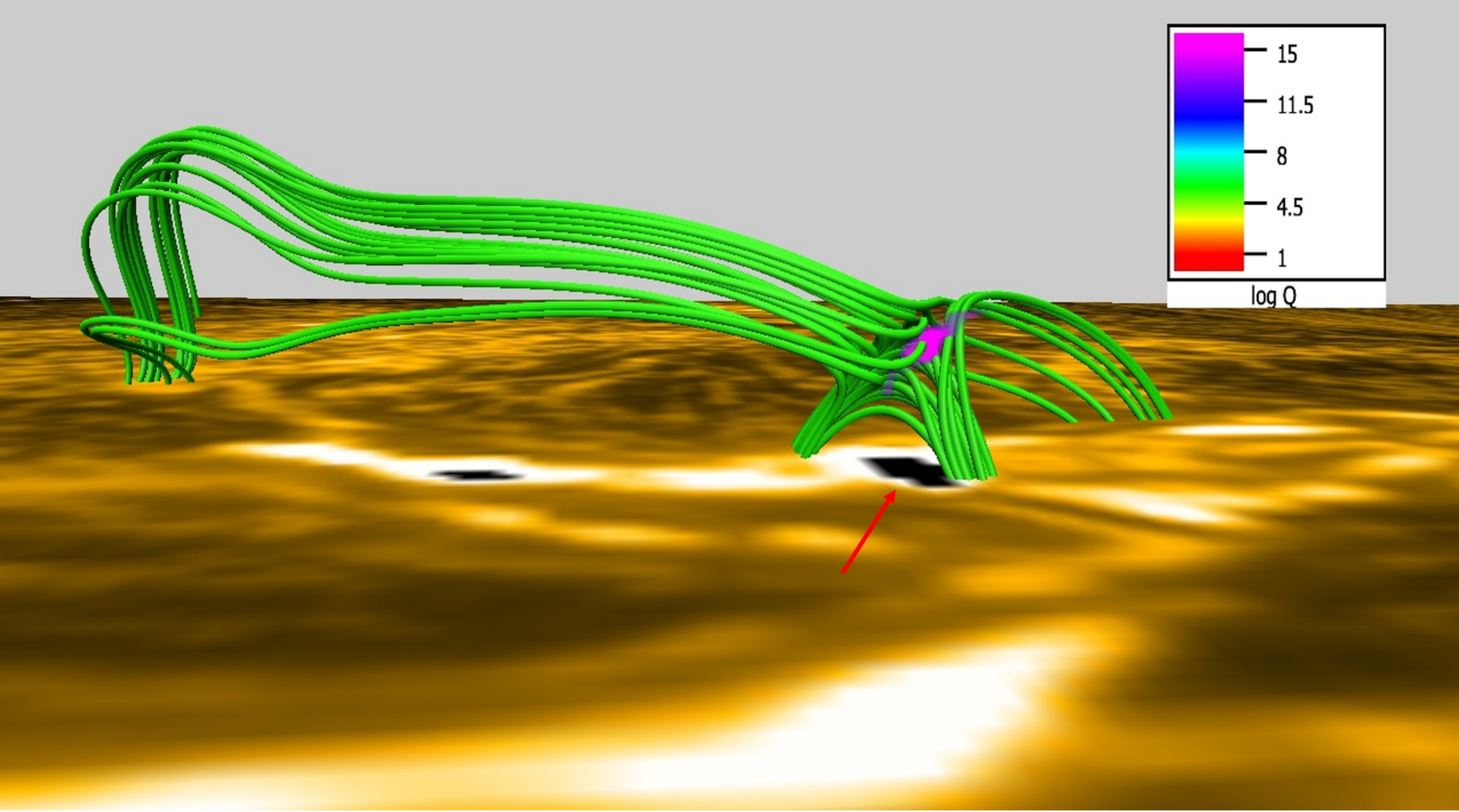}
\includegraphics[width=0.45\textwidth]{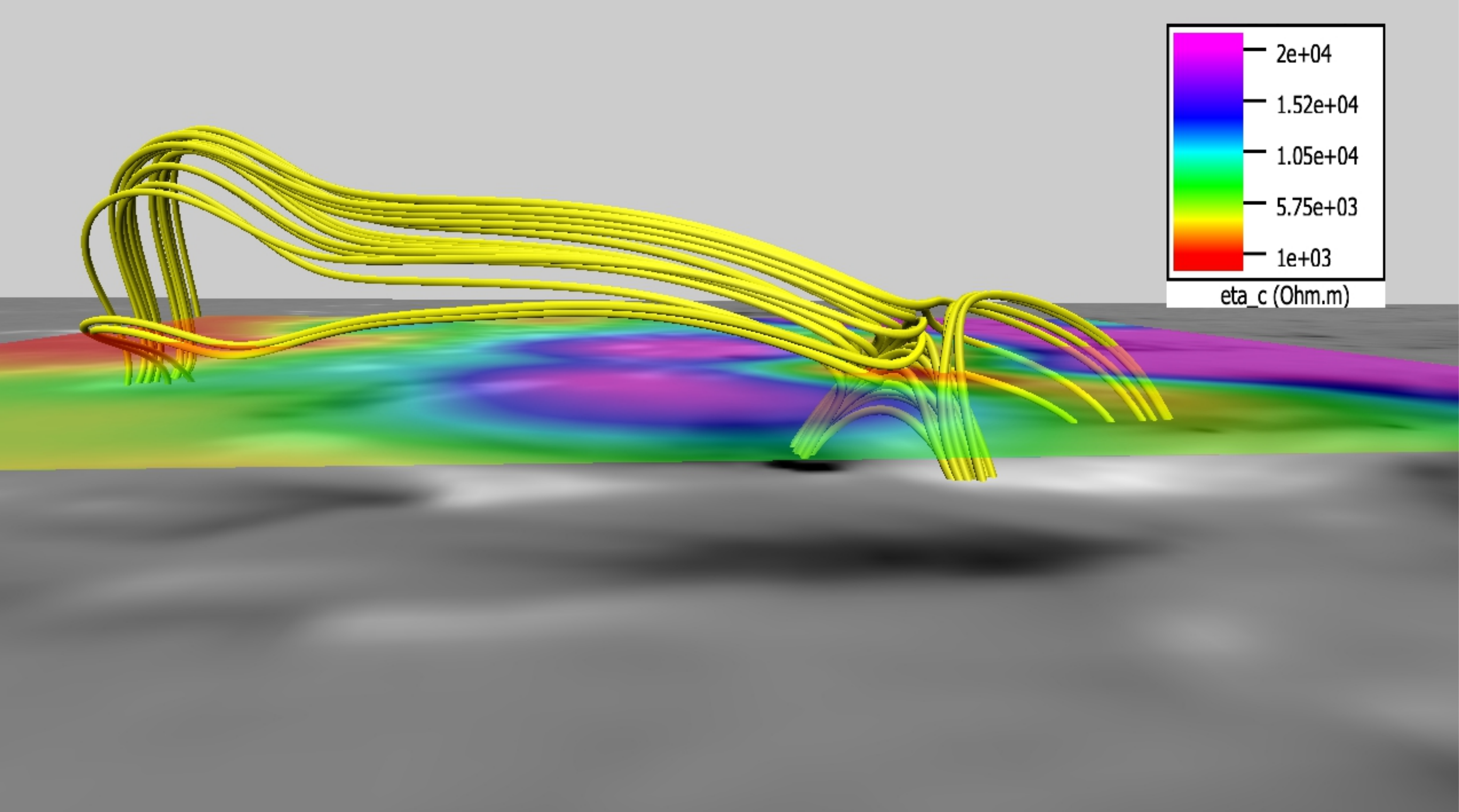}
\includegraphics[width=0.45\textwidth]{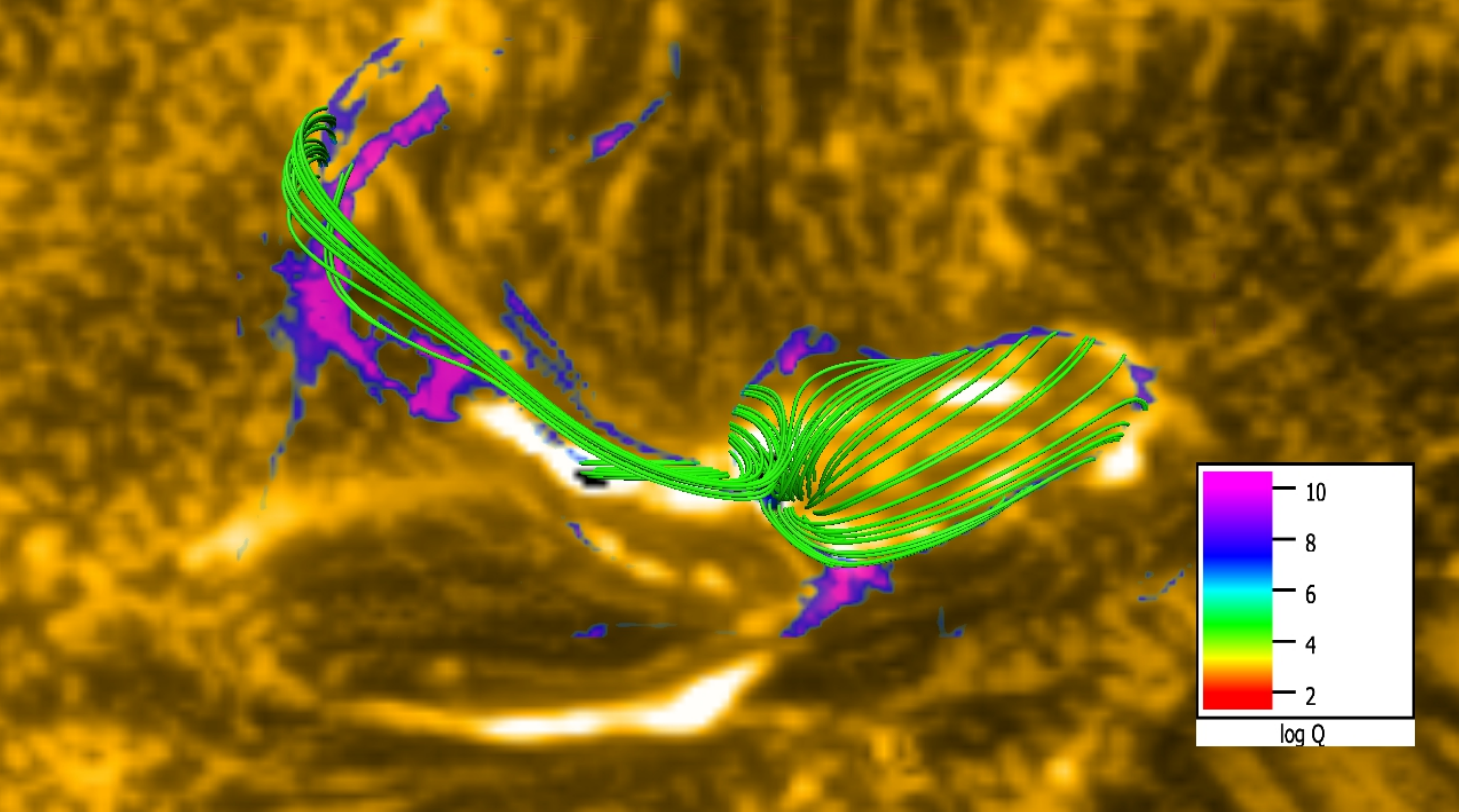}
\includegraphics[width=0.45\textwidth]{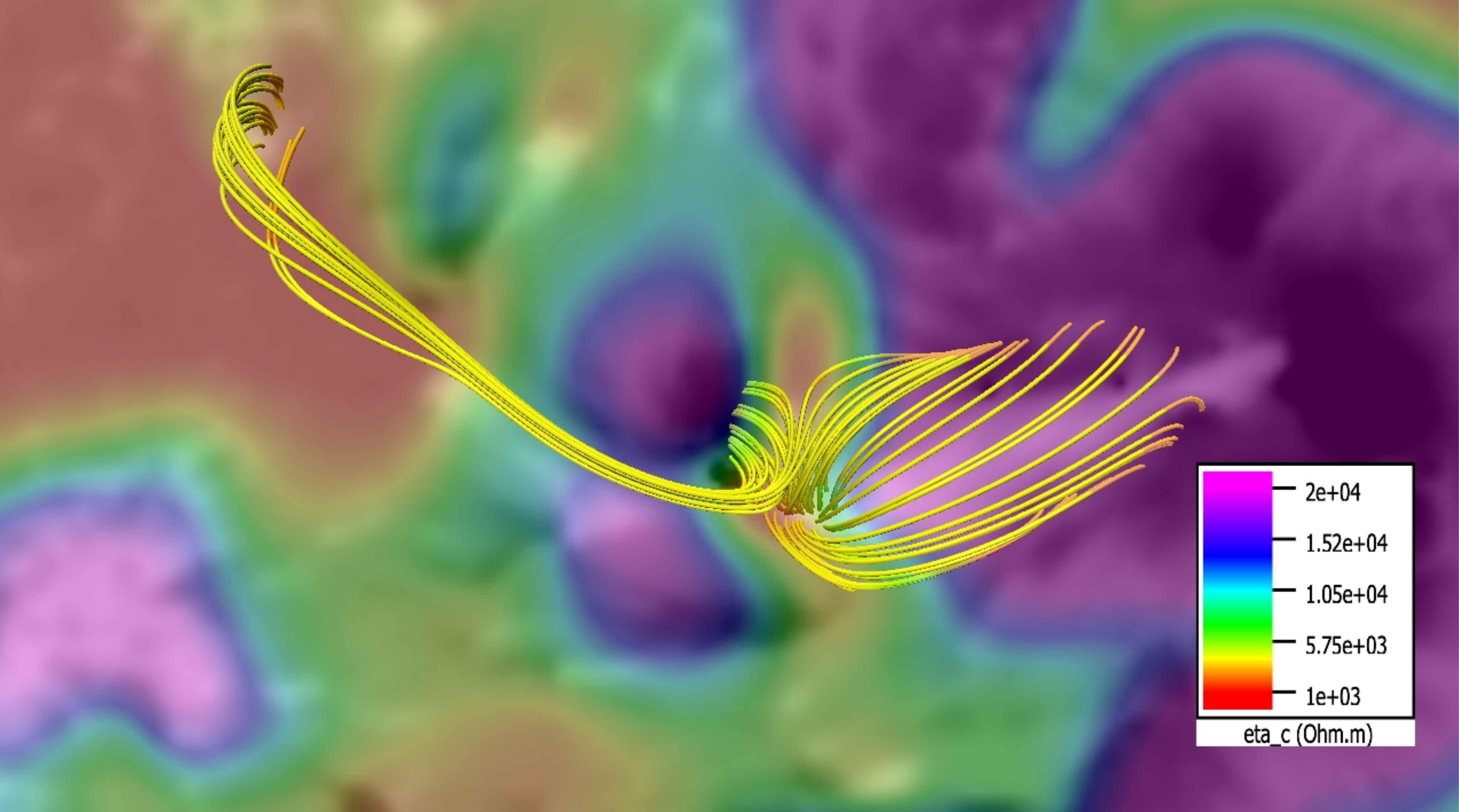}
\caption{(Left panel) Side and top views of a 3D null point with its corresponding spine-fan topology superimposed on an extreme-ultraviolet (EUV) channel 171 \AA~ image from \emph{SDO}/AIA at 2011-03-10T14:23:36 UT corresponding to a C2.0 flare (indicated by the red arrow). The squashing factor (log Q) \cite{Liu16} contours are shown at the location of the 3D null point (top), which is at a height of $\sim$1.9 Mm, and at the bottom boundary (bottom). (Right panel) Side and top views of the magnetic field configuration superimposed on an HMI magnetogram showing AR11166 at 2011-03-10T14:24 UT and the Cowling resistivity distribution ($\Omega$.m) just below the null point. The C2.0 flare location is at (326,255) arcsec or (N15.34,W20.46) degrees.}
\label{fig5}
\end{figure}

In Figure~\ref{fig5}, we show such a 3D null point configuration at 2011-03-10T14:23:36 UT resulting in the C2.0 flare (see Figure~\ref{fig5} caption for details) emerging from AR11166 with its location being in a region where the Cowling resistivity is dominant.

On following~\cite{VL99} that examines the role of ambipolar diffusion in Sweet-Parker reconnection, we find a normalized magnetic reconnection rate of 0.12. This value is in agreement with~\cite{Xue16} and the references therein.

We refer the interested reader to~\cite{Yalim20} for more details.

\section{Conclusions}
\label{conc}
We have developed a data-driven MHD model for the weakly-ionized chromosphere. In this paper, we presented an overview of this model and a benchmark case result which validated the local simulation setup. The partial ionization effects in our model equations are introduced by the Cowling resistivity which also poses numerical challenges such as the restriction on the stability condition and hence the convergence speed that we needed to address.

We also discussed the effects of the Cowling resistivity in the weakly-ionized chromosphere during the evolution of an AR and on flare formation associated with magnetic reconnection in the chromosphere.

We analyzed the evolution of AR11166. The Cowling resistivity is found to be orders of magnitude larger than the Coulomb resistivity, especially between 1-2 Mm height in the chromosphere. It has a significant effect on the chromospheric heating via frictional Joule heating due to current dissipation perpendicular to the magnetic field. The time-dependent evolution of Cowling resistivity gives an indication about the evolution of the AR as well since it follows the AR structure quite closely due to its strong dependence on the magnetic field strength. 

We also analyzed the effect of Cowling resistivity on the formation of a C2.0 flare that emerged from AR11166. The Cowling resistivity can have an effect on flare formation for a low-lying 3D null point configuration that occurs at a height of $\sim$1.9 Mm where the Cowling resistivity has its largest value. We obtained a normalized magnetic reconnection rate of 0.12 which is in agreement with~\cite{Xue16} and the references therein. We also found a good match between the AIA brightening with the log Q contours and the location of the null point inferred from the extrapolated magnetic field topology.

In a future work, we will perform a numerical simulation of the evolution of AR11166 using our model and compare with the results in this paper.

We acknowledge support from the NSF EPSCoR RII-Track-1 Cooperative Agreement OIA-1655280. Any opinions, findings, and conclusions or recommendations expressed in this material are those of the author(s) and do not necessarily reflect the views of the National Science Foundation. M.S.Y. and N.P. acknowledge partial support from NASA LWS grant 80NSSC19K0075. A.P. and Q.H. acknowledge partial support from NASA grant 80NSSC17K0016 and NSF award AGS-1650854.  

The HMI and AIA data have been used courtesy of NASA/\emph{SDO}, and HMI and AIA science teams.


\end{document}